\documentclass[12pt,letterpaper]{article}
\usepackage{amsmath,amssymb,array,calc,rotating,epsfig,psfrag, amscd}

\usepackage[left=2cm,top=1cm,right=3cm,nohead]{geometry}

\begin{document}


\newcommand{\nc}{\newcommand}

\nc{\ra}{\rightarrow} 
\nc{\lra}{\leftrightarrow} 
\nc{\Ra}{\Rightarrow} 
\nc{\LRa}{\Leftightarrow} 
\nc{\blp}{{\big (}}
\nc{\brp}{{\big )}}
\nc{\Blp}{{\Big (}}
\nc{\Brp}{{\Big )}}
\nc{\bglp}{{\bigg (}}
\nc{\bgrp}{{\bigg )}}
\nc{\Bglp}{{\Bigg (}}
\nc{\Bgrp}{{\Bigg )}}
\nc{\slb}{{\rm [}}
\nc{\srb}{{\rm ]}}
\def\al{\alpha}
\def\bet{\beta}
\def\ch{\chi}
\def\eps{\epsilon}
\nc{\veps}{\varepsilon}
\def\gam{\gamma}
\def\i{\iota}
\def\kap{\kappa}
\def\lam{\lambda}
\def\om{\omega}
\nc{\vphi}{\varphi}
\def\tha{\theta}
\def\sig{\sigma}
\def\ups{\upsilon}
\def\Gam{\Gamma}
\def\Lam{\Lambda}
\def\Om{\Omega}
\def\Sig{\Sigma}
\def\Tha{\Theta}
\def\Ups{\Upsilon}
\nc{\bea}{\begin{eqnarray}}
\nc{\eea}{\end{eqnarray}}
\nc{\be}{\begin{equation}}
\nc{\ee}{\end{equation}}
\nc{\cA}{{\cal A}}
\nc{\cB}{ \cal B}
\def\cC{{\cal C}}
\def\cD{{\cal D}}
\def\cE{{\cal E}}
\nc{\cF}{{\cal F}}
\nc{\cG}{{\cal G}}
\def\cH{{\cal H}}
\def\cI{{\cal I}}
\def\cJ{{\cal J}}
\def\cK{{\cal K}}
\nc{\cL}{{\cal L}}
\nc{\cM}{{\cal M}}
\def\N{{\cal N}}
\def\cN{{\cal N}}
\def\cO{{\cal O}}
\def\cP{{\cal P}}
\nc{\cQ}{{\cal Q}}
\nc{\cR}{{\cal R}}
\def\cS{{\cal S}}
\def\cT{{\cal T}}
\def\cV{{\cal V}}
\def\cV{{\cal V}}
\def\cW{{\cal W}}
\def\cX{{\cal X}}
\def\cY{{\cal Y}}
\def\cZ{{\cal Z}}
\nc{\cQd}{\cQ^{\dagger}}
\nc{\cRd}{\cR^{\dagger}}
\nc{\BB}{{\mathbb B}}
\nc{\CC}{{\mathbb C}}
\nc{\DD}{{\mathbb D}}
\nc{\EE}{{\mathbb E}}
\nc{\FF}{{\mathbb F}}
\nc{\GG}{{\mathbb G}}
\nc{\HH}{{\mathbb H}}
\nc{\JJ}{{\mathbb J}}
\nc{\RR}{{\mathbb R}}
\nc{\PP}{{\mathbb P}}
\nc{\QQ}{{\mathbb Q}}
\nc{\ZZ}{{\mathbb Z}}
\nc{\calone}{{\mathbb 1}}
\newcommand{\oo}{\frac{1}}
\nc{\half}{\frac{1}{2}}
\nc{\qrt}{\frac{1}{4}}
\nc{\del}{\partial}
\def\pder#1#2{{\frac{\partial{#1}}{\partial{#2}}}}
\nc{\delbar}{\bar\partial}
\nc{\Spin}{\operatorname{Spin}}
\nc{\SO}{\operatorname{SO}}
\renewcommand{\O}{\operatorname{O}}
\nc{\Sp}{{\rm Sp}}
\nc{\com}[2]{{ \left[ #1, #2 \right] }}
\nc{\acom}[2]{{ \left\{ #1, #2 \right\} }}
\nc{\rr}{\rightarrow}
\nc{\p}{\partial}
\nc{\LT}{{\LL_\T}}
\nc{\Tr}{{\rm Tr}}
\nc{\tr}{{\rm tr}}
\def\com#1#2{{ \left[ #1, #2 \right] }}
\def\acom#1#2{{ \left\{ #1, #2 \right\} }}
\nc{\Adag}{A^{\dagger}}
\nc{\AdagI}{A^{\dagger I}}
\nc{\AdagJ}{A^{\dagger J}}
\nc{\AdagK}{A^{\dagger K}}
\nc{\AdagL}{A^{\dagger L}}
\nc{\AdagM}{A^{\dagger M}}
\nc{\Bdag}{B^{\dagger}}
\nc{\BdagI}{B^{\dagger}_I}
\nc{\BdagJ}{B^{\dagger}_J}
\nc{\BdagK}{B^{\dagger}_K}
\nc{\BdagL}{B^{\dagger}_L}
\nc{\BdagM}{B^{\dagger}_M}
\nc{\Cdag}{C^{\dagger}}
\nc{\CdagI}{C^{\dagger I}}
\nc{\CdagJ}{C^{\dagger J}}
\nc{\CdagK}{C^{\dagger K}}
\nc{\Ddag}{D^{\dagger}}
\nc{\DdagI}{D^{\dagger I}}
\nc{\DdagJ}{D^{\dagger J}}
\nc{\DdagK}{D^{\dagger K}}
\nc{\ttha}{\tilde{\theta}}
\nc{\tphi}{\tilde{\phi}}
\nc{\tsig}{\tilde{\sig}}
\nc{\tom}{\tilde{\om}}
\nc{\tlam}{\tilde{\lam}}
\nc{\tSig}{\widetilde{\Sig}}
\nc{\tPhi}{\tilde{\Phi}}
\nc{\tPhibar}{\ol{\tPhi}}
\nc{\tPi}{\tilde{\Pi}}
\nc{\tpsi}{\tilde{\psi}}
\nc{\tPsi}{\tilde{\Psi}}
\nc{\tgam}{\tilde{\gam}}
\nc{\tGam}{\tilde{\Gam}}
\nc{\tb}{\tilde b}
\nc{\tc}{\tilde c}
\nc{\te}{\tilde e}
\nc{\tf}{\tilde f}
\nc{\tg}{\tilde g}
\nc{\tj}{\tilde j}
\nc{\tp}{\widetilde{p}}
\nc{\tq}{\widetilde{q}}
\nc{\ts}{{\tilde s}}
\nc{\tu}{{\tilde u}}
\nc{\tv}{{\tilde v}}
\nc{\tw}{{\tilde w}}
\nc{\tx}{{\tilde x}}
\nc{\ty}{{\tilde y}}
\nc{\tz}{\tilde z}
\nc{\tA}{{\tilde A}}
\nc{\tAbar}{{\ol \tA}}
\nc{\tD}{{\tilde D}}
\nc{\tE}{{\tilde E}}
\nc{\tG}{{\tilde G}}
\nc{\tH}{{\tilde H}}
\nc{\tJ}{{\tilde J}}
\nc{\tJbar}{{\ol {\tilde J}}}
\nc{\tM}{{\tilde M}}
\nc{\tN}{{\tilde N}}
\nc{\tP}{{\tilde P}}
\nc{\tQ}{{\tilde Q}}
\nc{\tR}{{\tilde R}}
\nc{\tS}{\tilde{S}}
\nc{\tF}{\tilde{{\cal F}}}
\nc{\tX}{\widetilde{X}}
\nc{\tcZ}{\tilde{\cZ}}
\nc{\tcZbar}{\ol{\tcZ}}
\nc{\hb}{\hat b}
\nc{\hc}{\hat c}
\nc{\hd}{\hat d}
\nc{\he}{\hat e}
\nc{\hf}{\hat f}
\nc{\hg}{\hat g}
\nc{\hh}{\hat h}
\nc{\hp}{\hat p}
\nc{\hv}{\hat v}
\nc{\hw}{\hat w}
\nc{\hx}{\hat x}
\nc{\hy}{\hat y}
\nc{\hz}{\hat z}
\nc{\hA}{\widehat{A}}
\nc{\hE}{\widehat{E}}
\nc{\hF}{\widehat{F}}
\nc{\hH}{\widehat{H}}
\nc{\hJ}{\widehat{J}}
\nc{\tK}{\widetilde{K}}
\nc{\hM}{\widehat M}
\nc{\hV}{\widehat V}
\nc{\hcV}{\widehat \cV}

\nc{\ha}{\widehat \alpha}
\nc{\hphi}{\hat{\phi}}
\nc{\hpsi}{\hat{\psi}}
\nc{\hgam}{\hat{\gam}}
\nc{\hPhi}{\hat{\Phi}}
\nc{\hPsi}{\hat{\Psi}}
\nc{\hGam}{\hat{\Gam}}

\nc{\w}{\wedge}


\nc{\vb}{\vec b}
\nc{\vc}{\vec c}
\nc{\vd}{\vec d}
\nc{\ve}{\vec e}
\nc{\vf}{\vec f}
\nc{\vg}{\vec g}
\nc{\vh}{\vec h}
\nc{\vp}{\vec p}
\nc{\vq}{\vec q}
\nc{\vr}{\vec r}
\nc{\vs}{\vec s}
\nc{\vv}{\vec v}
\nc{\vw}{\vec w}
\nc{\vx}{\vec x}
\nc{\vy}{\vec y}
\nc{\vz}{\vec z}

\nc{\ol}{\overline}
\nc{\abar}{\ol{a}}
\nc{\bbar}{\ol{b}}
\nc{\cbar}{\ol{c}}
\nc{\dbar}{\ol{d}}
\nc{\ebar}{\ol{e}}
\nc{\ibar}{\ol{\imath}}
\nc{\jbar}{\ol{\jmath}}
\nc{\kbar}{\ol{k}}
\nc{\lbar}{\ol{l}}
\nc{\mbar}{\ol{m}}
\nc{\nbar}{\ol{n}}
\nc{\pbar}{\ol{p}}
\nc{\qbar}{\ol{q}}
\nc{\ubar}{\ol{u}}
\nc{\vbar}{\ol{v}}
\nc{\wbar}{\ol{w}}
\nc{\xbar}{\ol{x}}
\nc{\ybar}{\ol{y}}
\nc{\zbar}{\ol{z}}

\nc{\Abar}{\ol{A}}
\nc{\Bbar}{\ol{B}}
\nc{\Cbar}{\ol{C}}
\nc{\Dbar}{\ol{D}}
\nc{\Ebar}{\ol{E}}
\nc{\Jbar}{\ol{J}}
\nc{\Kbar}{\ol{K}}
\nc{\Lbar}{\ol{L}}
\nc{\Pbar}{\ol{P}}
\nc{\Qbar}{\ol{Q}}
\nc{\Rbar}{\ol{R}}
\nc{\Sbar}{\ol{S}}
\nc{\Tbar}{\ol{T}}
\nc{\Ubar}{\ol{U}}
\nc{\Vbar}{\ol{V}}
\nc{\Wbar}{\ol{W}}
\nc{\Xbar}{{\overline X}}
\nc{\Ybar}{{\overline Y}}
\nc{\Zbar}{{\overline Z}}
\nc{\cZbar}{{\overline \cZ}}

\nc{\epsbar}{\ol{\epsilon}}
\nc{\lambar}{\ol{\lambda}}
\nc{\psibar}{\ol{\psi}}
\nc{\Psibar}{\ol{\Psi}}
\nc{\phibar}{\ol{\phi}}
\nc{\Phibar}{\ol{\Phi}}
\nc{\chibar}{\ol{\chi}}
\nc{\mubar}{\ol{\mu}}
\nc{\nubar}{\ol{\nu}}
\nc{\rhobar}{\ol{\rho}}
\nc{\ombar}{\ol{\om}}
\nc{\Ombar}{\ol{\Om}}

\nc{\gdot}{\dot{g}}
\nc{\xdot}{\dot{x}}
\nc{\ydot}{\dot{y}}
\nc{\sinp}{s_{\phi}}
\nc{\cosp}{c_{\phi}}
\nc{\tanp}{t_{\phi}}
\nc{\spone}{s_{\phi_1}}
\nc{\cpone}{c_{\phi_1}}
\nc{\tpone}{t_{\phi_1}}
\nc{\sptwo}{s_{\phi_2}}
\nc{\cptwo}{c_{\phi_2}}
\nc{\tptwo}{t_{\phi_2}}
\nc{\spth}{s_{\phi_3}}
\nc{\cpth}{c_{\phi_3}}
\nc{\tpth}{t_{\phi_3}}

\nc{\csch}{{\rm csch}}

\nc{\bah}{{\mathbf {\hat{A}}}}
\nc{\bX}{{\mathbf X}}
\nc{\ba}{{\bf a}}
\nc{\bb}{{\bf b}}
\nc{\bc}{{\bf c}}
\nc{\bd}{{\bf d}}
\nc{\bg}{{\bf g}}
\nc{\bk}{{\bf k}}
\nc{\bl}{{\bf l}}
\nc{\bm}{{\bf m}}
\nc{\bn}{{\bf n}}
\nc{\bo}{{\bf o}}
\nc{\bp}{{\bf p}}
\nc{\bq}{{\bf q}}
\nc{\br}{{\bf r}}
\nc{\bs}{{\bf s}}
\nc{\bt}{{\bf t}}
\nc{\bu}{{\bf u}}
\nc{\bv}{{\bf v}}
\nc{\bw}{{\bf w}}
\nc{\bx}{{\bf x}}
\nc{\by}{{\bf y}}
\nc{\bz}{{\bf z}}
\nc{\bom}{{\bf \om}}
\nc{\bombar}{{\mathbf \ombar}}
\nc{\bPhi}{{\bf \Phi}}

\nc{\rma}{{\rm a}}
\nc{\rmb}{{\rm b}}
\nc{\rmc}{{\rm c}}
\nc{\rmd}{{\rm d}}
\nc{\rmg}{{\rm g}}
\nc{\rk}{{\rm k}}
\nc{\rml}{{\rm l}}
\nc{\rmm}{{\rm m}}
\nc{\rmn}{{\rm n}}
\nc{\rmo}{{\rm o}}
\nc{\rmp}{{\rm p}}
\nc{\rmq}{{\rm q}}
\nc{\rmr}{{\rm r}}
\nc{\rms}{{\rm s}}
\nc{\rmt}{{\rm t}}
\nc{\rmu}{{\rm u}}
\nc{\rmv}{{\rm v}}
\nc{\rmw}{{\rm w}}
\nc{\rmx}{{\rm x}}
\nc{\rmy}{{\rm y}}
\nc{\rmz}{{\rm z}}

\nc{\dal}{\dot{\al}}
\nc{\thadot}{\dot{\tha}}
\nc{\thab}{\bar{\theta}}
\nc{\thal}{\theta^{\al}}
\nc{\thdal}{\bar{\theta}^{\dal}}

\nc{\thsigthm}{\tha \sigma^m \thab}
\nc{\thsigthn}{\tha \sigma^n \thab}

\nc{\Dal}{D_{\al}}
\nc{\Ddal}{\bar{D}_{\dal}}
\nc{\CDal}{{\cal D}_{\al}}
\nc{\CDdal}{\bar{\cal D}_{\dal}}

\nc{\eq}[1]{(\ref{#1})}
\nc{\non}{\nonumber}
\def\eg{{\it e.g.}}
\def\ie{{i.e.}}
\nc{\equ}{{\rm eq}}
\def\Im{{\rm Im ~}}
\def\Re{{\rm Re ~}}
\nc{\AdS}{{\rm AdS}}
\nc{\vol}{{\rm vol}}
\nc{\Ainf}{A_{\infty}}
\nc{\End}{{\rm End}}
\nc{\Ext}{{\rm Ext}}
\nc{\Hom}{{\rm Hom}}
\nc{\IIB}{{\rm IIB}}
\nc{\Ad}{{\rm Ad}}
\nc{\Dslash}{\ensuremath \raisebox{0.025cm}{\slash}\hspace{-0.32cm} D}
\nc{\cDslash}{\ensuremath \raisebox{0.025cm}{\slash}\hspace{-0.32cm} \cD}
\nc{\no}{\!:\!\!}
\nc{\ointdz}{\oint\frac{dz}{2\pi i}}
\nc{\ointdzone}{\oint\frac{dz_1}{2\pi i}}
\nc{\ointdztwo}{\oint\frac{dz_2}{2\pi i}}
\nc{\ointdzb}{\oint\frac{d\zbar}{2\pi i}}
\nc{\ointdzbone}{\oint\frac{d\zbar_1}{2\pi i}}
\nc{\ointdzbtwo}{\oint\frac{d\zbar_2}{2\pi i}}
\nc{\dz}{\frac{dz}{2\pi i}}
\nc{\dzb}{\frac{d\zbar}{2\pi i}}
\nc{\bpm}{\begin{pmatrix}}
\nc{\epm}{\end{pmatrix}}
 \nc{\bitem}{\begin{itemize}}
 \nc{\eitem}{\end{itemize}}
 
 \newtheorem{Thm}{Theorem}[section]
\newtheorem{Prop}[Thm]{Proposition}
\newtheorem{Lem}[Thm]{Lemma}
\newtheorem{Cor}[Thm]{Corollary}
\newtheorem{Conj}[Thm]{Conjecture}
\newtheorem{Def}[Thm]{Definition}
\newtheorem{Exa}[Thm]{Example}
\newtheorem{Cond}[Thm]{Condition}
\newtheorem{Ass}{Assumption}
\newtheorem{lem}[Thm]{Lemma}
\newtheorem{cor}[Thm]{Corollary}
\newtheorem{pro}[Thm]{Proposition}
\newtheorem{ex}[Thm]{Example}
\newtheorem{remark}[Thm]{Remark}
\newtheorem{defi}[Thm]{Definition}

\begin{center}
\vskip 2 cm

{\Large \bf  Non-geometric Backgrounds  \\ and the First Order String Sigma Model} \\

\vskip 1.25 cm 
{ Nick Halmagyi\footnote{email address: halmagyi@lpt.ens.fr} }\\
\vskip 5mm
{\it Laboratoire de Physique Th\' eorique et
Hautes Energies,  CNRS UMR 7589 \\{\it Universit\' e
Pierre et Marie Curie, 4 Place Jussieu, 75252 Paris Cedex 05,
France \\
\vskip 5mm
 Laboratoire de
Physique Th\' eorique de l'\' Ecole Normale Sup\' erieure, \\
24 rue Lhomond, 75231 Paris, France }
\vskip 5mm
Institut de Physique Th\' eorique,
CEA/Saclay, CNRS-URA 2306,\\
Orme des Merisiers, F-91191 Gif sur Yvette, France}
\end{center}
\begin{abstract}
We study the first order form of the NS string sigma model allowing for worldsheet couplings corresponding on the target space to a bi-vector, a two-form and an inverse metric. Lifting the topological sector of this action to three dimensions produces several Wess-Zumino like terms which encode the bi-vector generalization of the Courant bracket. This bracket may be familiar to physicists through the $(H_{ijk},F_{ij}^{k},Q_i^{jk},R^{ijk})$ notation for non-geometric backgrounds introduced by Shelton-Taylor-Wecht. The {\it non-geometricity} of the string theory in encoded in the global properties of the bi-vector, when the bi-vector is a section then the string theory is geometric. Another interesting situation emerges when one considers membrane actions which are not equivalent to string theories on the boundary of the membrane. Such a situation arises when one attempts to describe the so-called $R$-space (the third T-dual of a $T^3$ with $H_3$ flux). This model appears to be, at least classically, described by a membrane sigma model, not a string theory. Examples of geometric backgrounds with bi-vector couplings and non-vanishing $Q$-coefficients are provided by gauged WZW models.

\end{abstract}

\newpage
\tableofcontents

\vskip 15mm
\section{Introduction}

Non-geometric string backgrounds were introduced by Hellerman-McGreevy-Williams \cite{Hellerman:2002ax} to encode the fact that string theory treats metric and fluxes on an equal footing. In the string sigma model, the metric and fluxes are a set of couplings which may be related to each other by duality transformations. As such the spacetime diffeomorphism group is generically enlarged to include other gauge transformations such as $B$-field shifts and $T$-duality. 

In perturbative string theory, the maximal duality group is realized by torus compactifications, on a $d$-dimensional torus the duality group is $O(d,d,\ZZ)$. Since this is distinctly a string theory effect (beyond the point particle limit of the string), it is highly desirable to understand non-geometric backgrounds from the string worldsheet. It has been realized for some time that a worldsheet sigma model which is covariant under this duality group should involve doubling the field space in some sense \cite{Tseytlin:1990va, Rocek:1991ps, Hull:2004in}. 

Another particularly enlightening approach to studying non-geometric backgrounds pioneered in \cite{Shelton:2005cf} is to study the four dimensional low energy action obtained from compactifying string theory on a $T^6$ allowing for certain non-geometric twists. This way they are lead to posit a realization of the Roytenberg bracket \cite{Roytenberg:math9910078, roytenberg-2002-61} in the low energy action, generalizing the results of \cite{Kaloper:1999yr} where geometric torus reductions were studied and similar results for the Courant bracket were found. The key conjecture of \cite{Shelton:2005cf} is that while certain terms in the bracket are not understood from a string theory point of view, they are nonetheless to be included to maintain invariance under the duality group. In this paper we are not concerned with the four dimensional low energy action at all but will draw inspiration from \cite{Shelton:2005cf}. We will make progress in the understanding of non-geometric string backgrounds from a worldsheet approach, with an view towards providing a structure which can be applied to examples more diverse than just non-geometric tori. 

Since the Roytenberg bracket is quite well understood in the symplectic geometry literature, our strategy is that by providing a worldsheet realization of this bracket we might transform non-geometric backgrounds into something more canonical in the math literature and thus make these backgrounds amenable to study with standard techniques. However this does not turn out to be quite the whole story. The results from math literature do not cover non-geometric backgrounds. The essential issue is that both the worldsheet description and the spacetime description hinge on the properties of a bi-vector $\Pi^{ij}$. In the math literature this bi-vector must be a global section of $\w^2T_X$ whereas in the string sigma model we can allow for the bi-vector to be locally defined and patched together with elements of the duality group. Of course this way of patching together a space is not a new concept when applied to the standard string worldsheet with metric and $B-$field couplings, this is the cornerstone of non-geometric backgrounds \cite{Hellerman:2002ax}.

In this paper we will not double the space of dynamical fields but instead we study a first order action which behaves nicely under the duality group by forming a covariant Hamiltonian, that is to say that we Legendre transform with respect to both space and time, arriving at
\be
\cS= \int_{\Sigma} \Blp \eta^{ab} \rmp_a\w* \rmp_b + e^a\w \rmp_a + B_{ab} e^a\w e^b + \Pi^{ab} \rmp_a \w \rmp_b    \Brp. \label{act1}
\ee
Lifting the topoloigical part of this action (that which doesn't depend on the worldsheet metric) to a membrane, in much the same way as Witten produced the Wess-Zumino term for the $B$-field \cite{Witten:1983ar}, produces a set of couplings parameterizing the Roytenberg bracket
\be
\cS=\int_{\Sig}\eta^{ab}p_a\w * p_b  +\int_M \Blp d\rmp_i  \psi^i + H_{ijk} \psi^i  \psi^j  \psi^k +F_{ij}^k\,   \psi^i  \psi^j \rmp_k  +Q^{ij}_k \psi^k  \rmp_i  \rmp_j  + R^{ijk}\, \rmp_i  \rmp_j  \rmp_k  \Brp  \label{act2}
\ee
where $\psi^i=dx^i+\Pi^{ik}p_k$. This action need also be only well-defined up to elements of the duality group. In all known examples, we find that the charges $(H,F,Q,R)$ transform into each other under the duality group whereas the bivector $\Pi$ (which appears explicitly through $\psi$) tranforms non-trivially.
We are thus left with the conclusion that $(H,F,Q,R)$ are good global objects whereas the bi-vector may be defined locally as long as one can patch it together using the duality group. So in this case the non-geometricity appears solely through $\Pi^{ij}$.

We find a formula expressing the charges $(H,F,Q,R)$ in terms of the worldsheet data \eq{HFQRframes}, reproducing a result found in \cite{Halmagyi:2008dr} from the current algebra on the string worldsheet. This is interesting in its own right, since it provides a way of computing the current algebra from the Lagrangian perspective. For example, we find that $R^{ijk}$ is
\be
R^{ijk}=[\Pi,\Pi]^{ijk}_S - (\w^3 \Pi H)^{ijk}
\ee
so that if one considers an action \eq{act2} with $R^{ijk}$ not of this form, then one can still interpret this as a membrane action which cannot be restricted to the boundary $\Sigma$. This possibility appears to be quite well appreciated in the geometry literature where arbitrary $(H,F,Q,R)$ are generally considered. As such one can interpret this as saying that these backgrounds are geometric but not locally string theories. 

The Courant bracket also appears in generalized complex geometry \cite{Hitchin:2004ut, Gualtieri:2003dx} as an integrability condition for a generalized almost complex structure. Further, it is now known that generalized complex geometry describes the most general $\cN=(2,2)$ supersymmetric string action \cite{Lindstrom:2004hi}. Recently Gualtieri has studied an equivalence between holomorphic Poisson geometry and generalized Kahler geometry \cite{gualtieri-2007}, it seems likely that this equivalence is related to the equivalence between the first order and the second order sigma model. Another interesting connection between the first order sigma model and generalized complex geometry is that the deformation space (belonging to a complex manifold $X$) of generalized complex structures \cite{Gualtieri:2003dx} is given by $H^p(X,\Lam^qT^{1,0})$ with $p+q=2$. This includes complex structue deforamtions, holomorphic $B$-fields and holomorphic bi-vectors. These are all possible marginal operators in the first order sigma model.

While many formal structures are now known regarding NS-flux backgrounds, there is still  surprisingly few examples. The canonical set of examples are (gauged) WZW models, particularly attractive since they are associated to a solvable conformal field theory. In this paper we will show that $G/H$ WZW models are canonical examples of $Q$-spaces.


\section{The First Order Action}

\subsection{The First Order Action as a Covariant Hamiltonian}

We consider a first order action, which can be obtained from the usual second order sigma model by forming a covariant Hamiltonian (performing a Legendre tranformation with respect to both space and time). The resulting action is
\be
\cS=-\frac{1}{4\pi \alpha'} \int_{\Sigma} \Blp \eta^{ab} \rmp_a\w* \rmp_b + e^a\w \rmp_a + B_{ab} e^a\w e^b + \Pi^{ab} \rmp_a \w \rmp_b    \Brp. \label{FOAction}
\ee
Here we have used $\rmp_a=p_{a\alpha}d\sig^{\alpha}$ to represent a worldsheet one-form with values in $T^*(M)$ and the frame is also a one form $e^a=e^{a}_{\ j} dx^j$ with $G_{ij} =e^a_{\ i} e^{b}_{\ j} \eta_{ab}$.  So in general we have tensors\footnote{here by $T_X$ we mean the tangent bundle not the holomorphic tangent bundle $T^{1,0}_X$, we do not assume that the target space has any complex structure.} in $H^p(X,\Lam^q T_X)$ where $p+q=2$. One easily recovers the more familiar Polyakov action 

\be
\cS_\cP = \frac{1}{4\pi \alpha'}\int d^2x \Blp g_{ij} dx^i \w*dx^j + \tilde{B}_{ij} dx^i \w dx^j \Brp \label{SPol}
\ee
by integrating out the one form $\rmp_a$. The relation between the fields in \eq{FOAction} and \eq{SPol} is 
\be
(G+\Pi)=(g+b)^{-1},\ \ \tilde{B}=B+b.
\ee

 From \eq{FOAction} one can easily pass to a more conventional Hamiltonian.  We first expand as
\bea
\cS &=& \int d\sig d\tau \Blp  (p_{i\sig}+B_{ik}\del x^k) \xdot^i - \half G^{ij} p_{i\sig} p_{j\sig} - \half G_{ij}( \del x^i +\Pi^{ik }p_{k\sig} )( \del x^j + \Pi^{jl} p_{l\sig})  \non \\
&&+\half G_{ij} \blp  p^{i}_{\tau} -(\del x^i +\Pi^{ik }p_{k\sig} ) \brp \blp p^j_{\tau} -( \del x^j + \Pi^{jl} p_{l\sig} )\brp \label{SHam}.
\eea
and then integrate out the massive auxiliary field $C^i=p^{i}_{\tau} -( \del_\sig x^i +\Pi^{ik }p_{k\sig})$. The resulting Hamiltonian is
\be
\cH= \half G^{ij} p_i p_j + \half G_{ij}( \del  x^i +\Pi^{ik }p_k )( \del  x^j + \Pi^{jl} p_l). \label{Ham}
\ee
From the first term in \eq{SHam} we see that the Poisson brackets are not the canonical ones but have been twisted \cite{Alekseev:2004np}
\bea
\{x^{i}(\sig),x^{j}(\sig') \}&=&0, \non \\
\{p_{i}(\sig),p_{j}(\sig') \}&=&-H_{ijk}\del x^k \delta(\sig-\sig'), \label{twbrs} \\
\{ x^{i}(\sig),p_{j}(\sig') \}&=&\delta^{i}_j\delta(\sig-\sig'). \non 
\eea

We can untwist the Poisson brackets by the transformation $p_i\ra p_i-B_{ik} \del x^k$:
\bea
\cH&\ra& \half G^{ij} (p_i -B_{ik} \del x^k ) (p_j -B_{jl} \del x^l ) \non \\
&& + \half G_{ij}( \del  x^i +\Pi^{ik }(p_k -B_{km} \del x^m  ))( \del  x^j + \Pi^{jl} (p_l-B_{ln} \del x^n))\label{Ham2}.
\eea
The Hamiltonians \eq{Ham} and \eq{Ham2} were considered in \cite{Halmagyi:2008dr} and here we see how they follow naturally from the simple action \eq{FOAction}. In that work we computed the worldsheet current algebra and interpreted the result as the bi-vector generalization of the Courant bracket. Here we will see the same structures emerge from a first order Lagrangian description.


\subsection{T-Duality and the First Order Action}

The first order action \eq{FOAction} has the obvious feature that it appears to incorporate fields taking values in both $T_X$ and $T^*_X$. As such one might expect that the action of $O(d,d)$ and T-duality in particular might be simple on this form of the action. We find however that the Hamiltonian \eq{Ham} is a more natural object in this regard where T-duality is performed in the canonical formalism as prescribed in \cite{Alvarez:1994wj}. This essentially means that $(\del x^i, p_i)$ transforms as an $O(d,d)$ vector and thus for example $\del y$ and $p_y$ are interchanged by T-duality along $y$. The physical interpretation is then straightforward, one is interchanging momentum modes with winding modes. Employing T-duality this way leads to a certain generalization of the Buscher rules to include $\Pi$-couplings.


\subsubsection{$\Pi=0$}
When $\Pi$ vanishes, one starts with a Hamiltonian
\bea
\cH&=& \half \oint d\sig \Blp g^{ij} (p_i + B_{ik} \del x^k)( p_j + B_{jl} \del x^l) + g_{ij} \del x^i \del x^j \Brp \non \\
&=& \half \oint d\sig\,   v^T E_B v \label{HamB}
\eea
where
\be
v=\bpm \del x \\ p \epm,\ \ \ E_B=\bpm g-Bg^{-1}B & -Bg^{-1} \\ g^{-1}B & g^{-1} \epm.
\ee
One can write this as
\bea
v^T E_{B} v &=& v^T \bpm 1& -B \\ 0 & 1\epm
 \bpm e^T &0\\0&(e^{-1})^T\epm
 \bpm e &0\\0&e^{-1}\epm
 \bpm 1 & 0 \\ B &1\epm v  \non \\
&=&\bpm \del x & p + B \del x\epm 
\bpm g&0\\0&g^{-1}\epm
\bpm \del x & p+B\del x \epm.
\eea
We abused notation somewhat by dropping the indices from the frames $e^a_{\ i}$ so that $e^T e = g$.

In general the action of $\cO \in O(d,d)$ is given by
\be
E\ra \cO^{T} E \cO \non
\ee
and so of course we see that the Hamiltonian \eq{HamB} is an $O(d,d)$ rotation by $
\cO_B=\bpm 1&0 \\ B &1\epm
$
of the Hamiltonian
\be
\cH=\half \oint d\sig v^T \bpm g&0\\0&g^{-1}\epm v. \non
\ee

It is instructive to recall T-duality in this context first. If we have a symmetry in a particular direction say $x^1$, then we may perform $T$-duality in this direction. This is tantamount to acting on the Hamiltonian with the involution $T_1\in O(d,d)$ where
\bea
&& T_{1,ii}=1, \ i\in\{2,\ldots d,d+2,\ldots 2d\}, \non \\
&& T_{1,1(d+1)}=T_{1,(d+1)1}=1, \\
&& T_{1,ij}=0\ o/w. \non
\eea
So acting as 
\be
v\ra T_1v \non
\ee
is equivalent to the following action on the Hamiltonian
\be
v^T T_1 E  T_1 v = v^T (T_1 O_B^T T_1) (T_1 \cO_g T_1) (T_1 \cO_B T_1) v
\ee
where $\cO_g=\bpm g &0 \\ 0 &g^{-1}\epm$. From this one could read of the Buscher rules \cite{Alvarez:1994wj}.


\subsubsection{$B=0$}

We now look at the case when the $B$-field vanishes but there is a non-vanishing $\Pi$-vector. Then we have
\be
\cH= \half \oint d\sig\,   v^T E_{\Pi} v \label{HamPi}
\ee
where
\bea
E_\Pi&=&\bpm 1 & 0 \\ -\Pi & 1  \epm \bpm g&  0 \\ 0&g^{-1}\epm
\bpm 1 & \Pi  \\ 0 & 1  \epm \non \\
&=& \bpm g &  g\Pi \\ -\Pi g & g^{-1} - \Pi g \Pi \epm 
 \eea
 Again, under $T$-duality along $x^1$, $E_\Pi$ transforms as 
 \be
 E_{\Pi} \ra T_1 E_{\Pi} T_1
 \ee
 and so we also have $\cO_{\Pi}\ra T_1 \cO_{\Pi} T_1$ and $\cO_{g}\ra T_1 \cO_g T_1$. 
 

 \subsubsection{$B\neq 0$ and $\Pi\neq 0$}
 Now in the case when both $\Pi$ and $B$ are nonvanishing we still can understand this as an $O(d,d)$ transformation of the pure metric background. We have that the Hamiltonian \eq{Ham2} is
 \be
 \cH= \half \oint d\sig\,   v^T E v \label{HamBPi}
 \ee
 where
 \bea
E_\Pi&=& 
\bpm 1& -B \\ 0 & 1\epm
\bpm 1 & 0 \\ -\Pi & 1  \epm
\bpm g&  0 \\ 0&g^{-1}\epm
\bpm 1 & \Pi  \\ 0 & 1  \epm
\bpm 1 & 0 \\ B &1\epm \non \\
&=& \bpm (1+B\Pi)g(1+\Pi B) - B g^{-1}B  & (1+ B\Pi) g\Pi -Bg^{-1} \\
\Pi g (1+ \Pi B) - g^{-1} B& g^{-1} - \Pi g \Pi \epm.
\eea
Once again $T$-duality acts by conjugation on $(\cO_B,\cO_\Pi,\cO_g)$.


 \subsection{The Courant Bracket}
\label{CBracket}
The Courant bracket on a manifold $X$ is a map $T_X\oplus T^*_X\ra T_X\oplus T^*_X$ (a very clear review is contained in \cite{Gualtieri:2003dx}) and the co-ordinate independant formula is
\be
[u+\alpha, v+\beta]_C= [u,v] + \cL_{u}\beta- \cL_{v} \alpha -\half d\blp\imath_u \alpha- \imath_v \beta \brp + H(u,v,.).
\ee
It is convenient to refer to the Courant bracket as the expression with $H=0$ and the twisted Courant bracket as the expression with $H\neq0$ but if we fail to make this distinction it will hopefully not confuse the reader.

In a basis of frames $e^a$ ($e_a$ are the dual basis of vector fields) we can write this in a way perhaps more familiar to physicists from works such as \cite{Kaloper:1999yr, Shelton:2005cf}
\bea
{\rm [}e_a,e_b{\rm ]}_C &=& f_{ab}^c e_c + H_{abc} e^k, \non \\
{\rm [}e_a, e^b{\rm }]_C&=& f_{ac}^b e^c. \non
\eea
It will prove useful to recall how one can derive the twisting of the Courant bracket by a two form $B$. One computes
\bea
&&{\rm [} e^B (u+\alpha), e^{B}(v+\beta) {\rm ]}_{\cC}-e^B {\rm [} u+\alpha,v+\beta {\rm ]}_{\cC} \non \\
&=& \imath_u \imath_v dB.
\eea

This bracket was derived from string theory in a nice paper by Alekseev and Strobl \cite{Alekseev:2004np}, they considered the current algebra of the usual string sigma model and found that
\be
\{ \cQ_{u,\alpha},\cQ_{v,\beta} \} = - \cQ_{[(u,\alpha),(v,\beta)]_C} ,
\ee
where 
\be
\cQ_{u,\alpha}=\oint d\sigma\blp u^ip_i + \alpha_j \del x^j  \brp .
\ee
In supersymmetric sigma models, the Courant bracket makes another appearance. The general off-shell $(2,2)$ sigma model with $H$-flux is known to contain chiral, twisted chiral and semi chiral superfields and the target space geometry is described by generalized complex geometry \cite{Gualtieri:2003dx, Lindstrom:2005zr}. Just as closure under the Lie bracket provides the condition for integrability of almost complex structures, closure under the Courant bracket provides the condition for integrability of generalized almost complex structures. It is most curious however that the Courant bracket can be found from the {\it bosonic} string sigma model, the structures studied in the current paper do not require supersymmetry and as such do not require the machinery of generalized complex geometry.


\subsection{The Roytenberg Bracket}
\label{RBracket}
The full Roytenberg bracket \cite{Roytenberg:math9910078, roytenberg-2002-61} is a generalization of the Courant bracket to include the remaining possible three tensors. Locally one can write it as\footnote{In the proper geometric setting this bracket is part of the defining data of an algebroid. With just $F\neq 0$ or $Q\neq0$ the relevant object is called a ${\it Lie\ algebroid}$,  with $(F,Q)\neq0$ a {\it Lie bi-algebroid}, with $(F,Q,R)\neq0$ a {\it quasi-Lie bialgebroid}, with $(H,F,Q)\neq0$ a {\it Lie quasi bialgebroid} and with  $(H,F,Q,R)\neq0$ a {\it proto lie-bialgebroid}. }
\bea
{\rm [}e_a,e_b{\rm ]}_{\cR} &=& F_{ab}^c e_c + H_{abc} e^c, \label{coeffs1}\\
{\rm [}e_a,e^b{\rm ]}_{\cR} &=& Q^{bc}_a e_c+F_{ac}^b e^c, \label{coeffs2}\\
{\rm [}e^a,e^b{\rm ]}_{\cR} &=& R^{abc}e_c+Q^{ab}_c e^c.  \label{coeffs3}
\eea

Some classes of this bracket can be obtained by twisting the Courant bracket by  bi-vector.
We first present a the Courant bracket twisted by a bi-vector, in much the same way as we computed the $H$-twisted Courant bracket. We wish to compute
\be
{\rm [}e^{\Pi}(u+\alpha),e^{\Pi}(v+\beta) {\rm ]}_{\cC} - e^{\Pi} [u+\alpha,v+\beta]_{\cC} 
\ee
and declare this the define the extra terms in the Roytenberg bracket. The first term gives
\be
{\rm [}e^{\Pi}(u+\alpha),e^{\Pi}(v+\beta) {\rm ]}_{\cC}= [u+\alpha, v+\beta]_{\cC} + [\alpha\Pi ,v]_{\cC} + [u,\beta\Pi ]_{\cC} + [\alpha \Pi ,\beta]_{\cC} + [\alpha, \beta \Pi]_{\cC} + [ \alpha \Pi , \beta \Pi ]_{\cC}
\ee
and the second term gives
\bea
e^{\Pi} [u+\alpha,v+\beta]_{\cC}&=& \Pi \Blp \cL_{u} \beta-\cL_v \alpha -\half d(\iota_u \beta - \iota_v \alpha) \Brp  + \Pi H(u,v,.) \non \\
&& -\w^2\Pi H (\alpha,.,v) + \w^2\Pi H (\beta,.,u)   +  \w^3\Pi H (\alpha,\beta,.)
\eea

This $\Pi$-twisted Courant bracket is clearly a bit more of a handful than the Courant bracket, many new terms appear. In co-ordinate independant notation it is given by
\bea
&&[u+\alpha,v+ \beta]_{\cR} \non \\
&&=[u,v] -H \Pi (u,v) \non \\
&& + \cL_{u}\beta - \cL_v \alpha -\half d (\iota_u \beta - \iota_v \alpha)  +  \Pi H  (\alpha, v,.) -  \Pi H( \beta,u,.)\non  \\
&&-[v,\alpha \Pi] + [u,\beta \Pi] +(\cL_v \alpha -\cL_{u}\beta  + \half d (\iota_u \beta - \iota_v \alpha))\Pi + \w^2\Pi H (\alpha,.,v) - \w^2\Pi H (\beta,.,u) \non \\
&& - [\alpha,\beta]_{\Pi} + \w^2\Pi H (\alpha,\beta,.)\non \\
&&+ H(u,v,.) + \Blp \half [\Pi,\Pi]_S - \w^3\Pi H\Brp(\alpha,\beta,.)\ . \label{royt}
\eea
This somewhat hefty expression may require a little navigation. Explicit expressions in a frame basis may help achieve this:
where the four tensors $(H,F,Q,R)$ are given by 
\bea
H_{abc}&=& (dB)_{abc}, \non \\
F_{ab}^c&=& f^c_{ab} - H_{abd} \Pi^{dc}, \label{HFQRframes} \\
Q_c^{ab}&=& \del_c\Pi^{ab} +f_{cd}^a \Pi^{db}-f_{cd}^b \Pi^{da}+ H_{cde} \Pi^{da} \Pi^{eb},  \non \\
R^{abc} &=& [\Pi,\Pi]^{abc}_S +\half( f_{de}^a \Pi^{bd} \Pi^{ce}-f_{de}^b \Pi^{ad} \Pi^{ce}-f_{de}^c \Pi^{bd} \Pi^{ae}) - \frac{1}{3!} H_{def} \Pi^{da} \Pi^{eb} \Pi^{fc} \non . 
\eea
The first line in \eq{royt} is represented by the coefficients $F_{ab}^c$  in \eq{coeffs1} and is a twisting of the Lie bracket due to the combined presence of $H$ and $\Pi$. The second line is the same coefficients $F_{ab}^c$ this time coming from \eq{coeffs2}. The third and fourth lines in \eq{royt} are the $Q$ coefficients from \eq{coeffs2} and \eq{coeffs3} respectively. The final line contains the $H$ and $R$ coefficients.

In \cite{Halmagyi:2008dr} we computed the algebra of charges, starting from the Hamiltonian \eq{Ham2} and found this bracket, much like the Courant bracket was found by Alekseev and Strobl \cite{Alekseev:2004np}. Specifically we found
\be
\{\cQ_{(u,\alpha)}, \cQ_{(v,\beta)} \}=- \cQ_{[(u,\alpha),(v,\beta)]_{\cR}}
\ee
where
\be
\cQ_{(u,\alpha)}= \oint d\sig \Blp u^i p_i + \alpha_i (\del x^i + \Pi^{ij} p_j)\Brp.
\ee

More generally however one has the possibility of brackets which do not come from $\Pi$-twisting. One of the points of emphasis from Roytenberg's work \cite{Roytenberg:math9910078, roytenberg-2002-61} is the possibility of defining $(H,F,Q,R)$ without reference to $B$ or $\Pi$. The interpretation of these more general brackets in string theory is an interesting challenge, which we address below. 

\subsection{Roytenberg Bracket as Generalized Wess-Zumino Terms}

One goal of the current work is to interpret the current algebra calculation of \cite{Halmagyi:2008dr} in a Lagrangian setting. This can be achieved by first dividing the action \eq{FOAction}  into topological terms and kinetic terms:
\bea
\cS&=&\cS_{kin}+\cS_{top}, \non \\
\cS_{kin}&=& -\frac{1}{4\pi \alpha' }\int_{\Sigma} g^{ij} \rmp_i\w * \rmp_j, \label{FOAkin}\\
\cS_{top}&=&\frac{1}{4\pi \alpha'} \int_{\Sigma} \Blp dx^i \w \rmp_i + B_{ij} dx^i \w dx^j + \Pi^{ij} \rmp_i \w \rmp_j    \Brp\label{FOAtop}
\eea
and then lifting $\cS_{top}$ to a membrane $M$ whose boundary is the string worldsheet $\del M=\Sigma$. This is as simple as taking the exterior derivative of  \eq{FOAtop}

\be
\cS_{top}\sim \int_M \Blp d\rmp_i ( dx^i +\Pi^{ij}\rmp_j)   + \del_i \Pi^{jk} dx^i  \rmp_j  \rmp_k  + \del_{[i}B_{jk]} dx^i  dx^j  dx^k \Brp  \label{muntw}
\ee
One can then make a certain field redefinition to twist the bracket
\be
\cS_{top}= \int_M \Blp d\rmp_i  \psi^i + H_{ijk} \psi^i  \psi^j  \psi^k +F_{ij}^k\,   \psi^i  \psi^j \rmp_k  +Q^{ij}_k \psi^k  \rmp_i  \rmp_j  + R^{ijk}\, \rmp_i  \rmp_j  \rmp_k  \Brp \label{membrane}
\ee
where we have used
\be
\psi^i=dx^i + \Pi^{ij} \rmp_j
\ee
and suppressed the wedge products. The tensors $(H,F,Q,R)$ are given by the expressions
\bea
H_{ijk}&=& \del_{[i} B_{jk]}, \non \\
F_{ij}^k&=& - H_{ijl} \Pi^{lk}, \label{HFQR} \\
Q_k^{ij}&=& \del_k\Pi^{ij} - H_{klm} \Pi^{li} \Pi^{mj},  \non \\
R^{ijk} &=& [\Pi,\Pi]^{ijk}_S - H_{lmn} \Pi^{li} \Pi^{mj} \Pi^{nk} \non . 
\eea
which are equivalent to \eq{HFQRframes}.

So we can now see the connection between the Hamiltonian and Lagrangian derivations of the Roytenberg bracket. The current algebra computation of \cite{Halmagyi:2008dr} starts from the Hamiltonian \eq{Ham} which is physically equivalent to the Lagrangian \eq{FOAction}. We then lift \eq{FOAction} to three dimensions \eq{membrane} and find the coefficients of the Roytenberg bracket as generalized Wess-Zumino couplings. 

The twisting procedure is a simple field redefinition and thus somewhat trivial. It might in fact be more reasonable to simply treat \eq{muntw} as defining the coefficients of the bracket however it is \eq{HFQR} which appear in the algebra of charges. On the other hand we will see below that gauged WZW models provide examples of backgrounds with $(H,F,Q)$ and there is an alternative approach of Getzler \cite{Getzler:1993fs} which suggests to use \eq{muntw} as defining the coefficients of the bracket.

A related calculation has in fact been performed before in a somewhat different context \cite{Park:2000au, Hofman:2002rv}. In that work, the authors define a topological membrane theory where the membrane in question would be the open membrane of M-theory. The boundary string theory would be something akin to a topological little string theory living on the five brane. In our setting, the membrane $M$ is not usually considered to be a physical object in the theory merely a mathematical tool for dealing with the global structure of the theory. 

Another key difference between the two scenarios is that we are concerned with the full dynamical sigma model, we have included \eq{FOAkin}. Disregarding the term \eq{FOAkin} one is left with what is sometimes called the WZ-Poisson model \cite{Klimcik:2001vg}. Since Hofman and Park were concerned only with the topological terms $\cS_{top}$ the theory had a large gauge symmetry which then naturally leads to the BV formalism. This gauge invariance the enforces  $R^{ijk}=0$ (see \cite{Hofman:2002rv} section 6.1). When adding the non-topological term \eq{FOAkin} this gauge symmetry is lost and this condition need not be enforced. It would be interesting to study extended supersymmetry in the full sigma model and see if one finds the Poisson condition. Further work on this topological model, often called the Courant sigma model can be found in \cite{Roytenberg:2006qz,Bonechi:2009kx, Hansen:2009zd}

\subsection{Quantum Aspects}

It is worth considering the beta-function computed from the model \eq{FOAction}.
We start by considering the flat background\footnote{we use the conventions of \cite{Polchinski:1998rq}: $ \eps_{01}=\eps^{10}=1,\ z=x^0+ix^1,\ \del=\half(\del_0-i\del_1),\ p=\half(p_0-ip_1)$}
\be
S=-\frac{1}{2 \pi \alpha'}\int d^2z \Blp  \delta^{ij}p_i \pbar_j + i\blp p_i  \delbar x^i- \pbar_i \del x^i \brp\Brp \label{FOAflat}
\ee
Clearly if we integrate out the $p_i$ fields we obtain the flat space second order action
\be
S=\frac{1}{2 \pi \alpha'}\int d^2z\, \delta_{ij} \del x^i \delbar x^j.
\ee
From the equations of motion of \eq{FOAflat}
\be
p=i\del x,\ \ \ \pbar = -i \delbar x,\ \ \del \pbar = \delbar p \label{eoms}
\ee
we deduce that
\be
\del \delbar x^i=0,\ \ \delbar p_i =0,\ \ \del \pbar_i =0
\ee
so there is decoupling between left and right movers: $(\del x^i, p_i)$ are holomorphic and $(\delbar x^i,\pbar_i)$ are anti-holomorphic. 

The two-point functions are 
\bea
x^i(z,\zbar) x^j(z',\zbar') &=&- \frac{\alpha'}{2} \delta^{ij} \ln |z-z'|^2,\ \ \  p_i (z) x^j(z',\zbar')= - \frac{i\alpha'}{2}\delta^{j}_{i}(z-z')^{-1}, \non \\
\pbar_i (\zbar) x^j(z',\zbar') &=&  \frac{i\alpha'}{2}\delta^{j}_{i}(\zbar-\zbar')^{-1},\ \ \ p_i(z) p_j(z') =  \frac{\alpha'}{2}\delta_{ij}(z-z')^{-2},  \\
\pbar_i (\zbar) \pbar_j (\zbar')&=&\frac{\alpha'}{2}\delta_{ij}(\zbar-\zbar')^{-2}  \non 
\eea
which is none too suprising since if we use the equations of motion for $(p,\pbar)$ then all these two-point functions agree with the standard two-point functions for a free boson.

The stress tensor is easy to read off
\be
T(z)=\frac{\delta^{ij}}{\alpha'}\no p_i p_j \no (z)
\ee
and we immediately see that $p_i$ has dimension $(1,0)$. The o.p.e. $T(z) \del x(z')$ is slightly more tricky since it initially doesn't seem $\del x$ is a primary field
\be
T(z) \del x(z') \sim -\frac{ip(z)}{(z-z')^2}
\ee
 however using \eq{eoms}  we see that $\del x(z)$ indeed has dimension $(1,0)$. 
 
 There are various classically marginal deformations to the action \eq{FOAflat} and the conditions for these operators  to be quantum mechanically marginal to first order simple. For example
  \be
 T(z) G^{ij}\no p_i \pbar_j\no(z',\zbar') \sim \frac{ \del_i G^{ij}\, \pbar_j(\zbar')}{(z-z')^3} +\frac{G^{ij}\no p_i \pbar_j\no(z',\zbar') }{(z-z')^2} +\frac{\del G^{ij} \no p_i \pbar_j\no(z',\zbar')}{z-z'}
 \ee
and so for this to have dimension $(1,1)$ we require that $\del_i G^{ij}=0$. Similarly we find by adding the operators
\be
i B_{ij} \del x \delbar x,\ \ i\Pi^{ij} p_i \pbar_j,\ \ i\mu^{i}_j (p_i\delbar x^j - \pbar_i \del x^j)
\ee
that we need
\be
\del_i \Pi^{ij}=0,\ \ \del^i B_{ij}=0,\ \ \del_i \mu^i_j=0, \ \ \del^j \mu^i_j=0. 
\ee

The beta function can now be found as usual by considering the two point function 
\be
\langle \Delta S(z,\zbar)  \Delta S(z',\zbar') \rangle \label{SS}
\ee
where 
\be
\Delta S= G^{ij} p_i \pbar j + i\Pi^{ij} p_i \pbar_j + iB_{ij} \del x^i \delbar x^j + i\mu^{i}_j(p_i\delbar x^j - \pbar_i \del x^j).
\ee
The key to performing this beta function calculation is to appreciate that all the possible couplings which appear with quadratic poles from the two point function \eq{SS} are equivalent to $\del x^i \delbar x^j$ by the equations of motion \eq{eoms}. This means that all possible contractions give quadratic poles.  The end result is somewhat disappointing, the beta-function calculation reduces to the beta-function calculation of the usual second order sigma model and thus produces the spacetime equations of motion in the same form. 

It is of course unavoidable that the conditions for conformal invariance of the first order and second order models are equivalent however there is the possibility that these conditions may be more revealing in terms of the couplings in the first order model. It is desirable to interpret the $(H,F,Q,R)$ coefficients as some independent flux quanta, as such one might expect them to contribute to the spacetime stress-tensor. It may be possible to rewrite the usual Einstein's equation in terms of $(G,\Pi)$ instead of $(g,b)$ in a such a way however the first order sigma model does not appear to provide this. 

A related but somewhat different first order action has been considered before \cite{Losev:2005pu, Nekrasov:2005wg}. This model has an action 
\be
S=\frac{1}{2\pi \alpha'}\int d^2z \Blp p_i \delbar x^i + \pbar_{\ibar} \del x^{\ibar} + G^{i\jbar} p_i \pbar_{\jbar} + i\Pi^{i\jbar} p_i \pbar_{\jbar} \Brp
\ee
where a complex structure has been selected on target space. One important difference when compared to our model is that $p_{\ibar}$ and $\pbar_i$ are absent which results in the peculiar relation
\be
g_{i\jbar} = B_{i\jbar},\ \ g_{\ibar j}=- B_{\ibar j} \label{gBijbar}
\ee
for the second order sigma model after integrating out the auxiliary $(p_{i}, \pbar_{\ibar})$ fields. When $(G,\Pi)\ra0$, this can be thought of as a singular limit of the usual NS sigma model where the metric and B-field are infinite.  The beta function for this model, computed by expanding around the point with $(G,\Pi)\ra 0$ gives interesting differential constraints of $(G,\Pi)$, quadratic in $(G,\Pi)$ which are equivalent to the usual beta function in the limit \eq{gBijbar}. This appears to be a special property of this background \eq{gBijbar} and in general, the beta function when expressed in terms of $(G,\Pi)$ has terms of arbitrarily high order in $(G,\Pi)$. 

It seems a little premature to impose a complex structure on the target space (and thus project out $(p_{\ibar},\pbar_i))$ when studying the bosonic string. Furthermore, looking forward to supersymmetrizing the model, without these fields $(p_{\ibar},\pbar_i)$ it is not possible to add a dimension $(1,1)$ operator corresponding to sections of $H^0(X,\Lam^2 T^{1,0})$ or $H^2(X,\Lam^{0}T^{1,0})$ since this corresponds to  terms like
\be
\int d^2z\,  \Pi^{ij} p_i \pbar_j,\ \ \ \ \int d^2 z B_{ij} \del x^i \delbar x^j
\ee
where $(i,j)$ are holomorphic indices. As explained earlier these fields appear in the deformation space of generalized complex structures. 


\subsection{Relation to Non-Geometric Backgrounds}

One main motivation for the current work was to study non-geometric string backgrounds, in particular to provide a worldsheet understanding of the algebras (\ref{coeffs1},\ref{coeffs2},\ref{coeffs3}) discussed in \cite{Shelton:2005cf, Dabholkar:2005ve} from the point of view of a four dimensional gravity effective action. The presence of arbitrary coefficients $(H,F,Q,R)$ was argued for by considering general properties of T-duality. We see here from the sigma model that there are two separate classes of these non-geometric backgrounds.

One possibility is that the action \eq{FOAction} is not well defined but one must use $O(d,d)$ valued transition functions instead of just diffeomorphisms and $B$-field shifts. From the action \eq{membrane} we see that to have the field strengths $(H,F,Q,R)$ globally well defined is still not quite enough since the bi-vector $\Pi$ itself appears explicitly in the action. For \eq{membrane} to be well defined globally one needs $\Pi$-to be a good section of $\w^2T$. If $\Pi$ is not a good section then one must be able to use $T$-duality to define it globally. The main point here is that original ideas of \cite{Hellerman:2002ax} can be translated purely into the properties of a bi-vector\footnote{a related argument was made in \cite{Grana:2008yw} by considering twists of $T\oplus T^*$ on the target space.}. To understand this better it would be nice to understand $T$-duality and its effect on the coefficients $(H,F,Q,R)$ at the level of the full sigma model. 

There is also another possibility. In the first order string sigma model one cannot realize the generic Roytenberg bracket but only the $\Pi$-twisted Courant bracket. The general Roytenberg bracket requires considering a membrane action which is not equivalent to a string theory on the boundary of the membrane\footnote{The non-topological term can also be lifted to the same membrane using Stokes theorem}. This membrane does not appear to be the membrane of M-theory since in M-theory one obtains the string by KK reduction, not by restriction to the boundary. This class of background appears to be related to what was termed "not locally geometric" in \cite{Shelton:2005cf}. Here we see it is in fact geometric but not really a string theory.


\section{Examples}

We now treat several examples. The canonical example of a string background with nontrivial $H$-flux is a WZW-model, in addition to the $H$-flux being non-zero it is also cohomologically non-trivial, there is no globally defined two-form $B$ such that $dB=H$. As a natural extension we see that {\it gauged} WZW models are canonical examples of $\Pi$-vector backgrounds. We find these models to have $H,F$ and $Q$ flux. We will first however revisit the beguiling example of a three torus with constant $H$ flux on its worldvolume and its associated T-duality frames.


\subsection{Three Torus with $H$-flux}
This well studied example appears to be particularly demonstrative regarding certain non-geometric structures. A three torus with uniform $H$-flux on it possesses three isometries and as such one should naively be able to $T$-dualize three times. The first $T$-duality is well understood but the final two provide canonical examples of non-geometric backgrounds.

\subsubsection{The Original Duality Frame: $H$-space}

This is the most clear duality frame
\bea
ds^2&=&dw^2+dx^2 + dy^2, \non \\
H&=&dw\w dx\w dy,\non \\
\Rightarrow B&=&N wdx\w dy. \non
\eea
\subsubsection{The Single $T$-dual: $F$-space}
After one $T$-duality along $y$ we have a purely metric background
\be
ds^2=dw^2 + dx^2 + (dy-Nwdx)^2.
\ee
\subsubsection{The Double $T$-dual: $Q$-space}
After another $T$-duality, using the Buscher rules we find
\bea
ds^2&=&dw^2+ \frac{1}{1+N^2 w^2} (dx^2+dy^2), \non \\
B&=&\frac{1}{(1+N^2 w^2)} dx\w dy.
\eea
As discussed in \cite{Halmagyi:2008dr} this is locally equivalent to the (closed-string) bi-vector background
\bea
ds^2&=& dw^2+dx^2+dy^2, \non \\
\Pi&=&Nw\del_x\w \del_y.
\eea
One can compute the $Q$-charge of this background using the Roytenberg bracket and one finds
\be
Q^{xy}_{w}=Q^{yw}_x=Q^{wx}_y=N.
\ee

This example is quite instructive regarding the general structure of \eq{membrane}. Here we have a non-trivial bi-vector background with a bi-vector which is only locally defined. As $w\ra w+1$ the bi-vector $\Pi$ shifts by an element of the duality group.
\be
\Pi\ra \Pi + N \del_x \w \del_y
\ee
So we see that the action \eq{membrane} is well defined only by using transitions functions on the target space more general than just diffeomorphisms however the field strength is well-defined.

\subsubsection{The Triple $T$-dual: $R$-space}
The third $T$-dual has for some time been somewhat mystifying. This is largely due to the explicit $w$-dependence in the $Q$-space background which appears to deny us the opportunity for a third $T$-duality, even though in the original duality frame we appeared to have three $T$-dualities available to us. One way to view the problem is that while the field strengths/charges,  have no $w$-dependance, the potentials (metric/$B$-field) do have $w$-dependance. This was our motivation to study the generalization of the Wess-Zumino terms in the membrane action \eq{membrane}, where the field strengths appear not the potentials. 

From this perspective we are naturally led to conjecture the following action as the third $T$-dual:
\be
\cS=\int_\Sigma \eta^{ab} \rmp_a\w * \rmp_b + \int_{M} \Blp d\rmp_i \w dx^i+ N \eps^{ijk} \rmp_i \w \rmp_j \w \rmp_k \Brp,\label{RSpace}
\ee
the metric is flat
\be
ds^2=dw^2+dx^2+dy^2.
\ee
This background is completely geometric, the radical nature of this action is that the second term cannot even locally be restricted to $\Sigma=\del M$ since $R$ is not of the form \eq{HFQR}. 

In general with cohomologically non-trivial $H$-flux, one cannot globally write the Wess-Zumino term as a coupling on the string worldsheet but one can however do so locally. Here we find that we cannot write this action \eq{RSpace} as a string theory even locally. This seems to leave two logical possibilities: backgrounds such as \eq{RSpace} are not part of string theory or alternatively the membrane action \eq{membrane} is more fundamental than the string action and these backgrounds are indeed part of our path integral, hopefully further work can clarify this issue.


\subsection{WZW Models}

It is also interesting to consider WZW models. In this case the action is
\be
I(g)= \frac{k}{8\pi}\int_{\Sigma=\del M} h_{ab} \sig^a \w * \sig^b+\Gam(g)
\ee
where
\be
\Gam(g)= \frac{k}{12\pi} \int_M f_{abc} \sig^a \w \sig^b\w \sig^c \non.
\ee
The invariant one forms $\sig^a$ are
\be
\sig^a = \Tr( \lambda^a g^{-1}dg)  \non 
\ee
and satisfy
\be
d\sig^a=\half f^{a}_{bc} \sig^b \w \sig^c \non
\ee
So the first order action for this model is
\bea
S_{WZW}&=& \frac{k}{4\pi}\int_{\Sigma} \Blp \sig^a\w \rmp_a - \half h^{ab} \rmp_a\w * \rmp_b\Brp  + \frac{k}{12\pi}\int_M f_{abc} \sig^a \w \sig^b \w\sig^c \non \\
&=&-\frac{k}{8 \pi}\int_{\Sigma} h^{ab} \rmp_a\w * \rmp_b  \non \\
&& + \frac{k}{12\pi}\int_{M} \Blp 3 d\rmp_a \w \sig^a+  f^a_{bc} \, \rmp_a\w \sig^b \w \sig^c + f_{abc} \sig^a \w\sig^b \w \sig^c \Brp \non.
\eea
So we see that the WZ terms are given by
\be
F^a_{bc}= f^{a}_{bc},\ \ \ \ H_{abc}=f_{abc}
\ee
which agrees with the Courant bracket for WZW models.

\subsection{$G/G$ Gauged WZW Models}

The gauging of WZW models is nicely explained in the appendix of \cite{Witten:1991mm}. The kinetic term is gauged with the standard minimal coupling but the WZ term requires a little more attention. One adds additional terms which are integrals over the string worldsheet $\Sigma$ but whose gauge variation cancels the gauge variation of the WZ term:

\be
I(g,A)= \frac{k}{8\pi}\int_\Sigma g^{-1}D_A g\w * g^{-1}D_A g + \Gam(g,A)
\ee
where
\be
\Gam(g,A)= \Gam(g) - \frac{k}{4\pi}\int_{\Sigma} \Blp A\w g^{-1}dg + A\w dg g^{-1} + Ag^{-1}Ag \Brp . 
\ee
In total one gets
\be
I(g,A)= I(g) - \frac{1}{2\pi} \int_\Sig \Blp A \delbar g\, g^{-1} - \Abar g^{-1} \del g + A \Abar - Ag\Abar g^{-1} \Brp.
\ee



The Hamiltonian for the $G/G$ model has been analysed before \cite{Alekseev:1995py} and here we  reinterpret this. The conventional Hamiltonian (just Legendre transforming time) for the $G/G$ model is the Hamiltonian of a twisted Poisson model, it is proportional to a constraint and is thus a topological sigma model. 

One can show that the action is
\bea
I(g,A)&=& \int_M \Tr (dg g^{-1})^3 +\int_\Sig d^2x \, \Tr \Blp p\, \gdot g^{-1} \non \\
&&-\Abar \blp g^{-1} pq -p + \frac{k}{4\pi}(g^{-1} g' + g' g^{-1}) \brp - pg'g^{-1} \non \\
&& -\frac{k}{8\pi} (A -\Abar- - \frac{4\pi}{k}p + g' g^{-1})^2 \Brp \label{GGHam2}
\eea
where
\bea
p&=& \frac{\delta \cL}{ \delta (\dot{g} g^{-1})} \non \\
&=& \gdot g^{-1} +A-g \Abar g^{-1} .
\eea
A somewhat subtle point is that we have not included the WZ term in $\cL$ since this can be absorbed covariantly into the Poisson brackets. The last line in \eq{GGHam2} is an auxiliary massive field, it can be integrated out. The term 
\be
\int_\Sig d^2x\, \Tr\,\Abar \blp g^{-1} pq -p + \frac{k}{4\pi}(g^{-1} g' + g' g^{-1}) \brp \non
\ee
is of the most interest, it is a set of Lagrange multipliers times a constraint. The final term 
\be
\int_\Sig d^2x \,pg'g^{-1} \label{extrapiece}
\ee 
appears to be awkward but in fact is set to zero by the constraint leaving us with
\be
\cH= \int d\sigma \Tr\,  \Abar \Blp g^{-1} pq -p + \frac{k}{4\pi}(g^{-1} g' + g' g^{-1})  \Brp. \label{GGHam}
\ee
This is similar to \cite{Witten:1993xi} where it was shown that the action of the $G/G$ model does in fact depend on the worldsheet metric but this dependance vanishes on-shell. As already mentioned we have absorbed the WZ term into twisted Poisson brackets for $p$ \eq{twbrs}.

Taking a closer look at the the constraint reveals the Poisson bivector,
\bea
&&\Tr\, \Abar\Blp g^{-1} pg -p + \frac{k}{4\pi}(g^{-1} g' + g' g^{-1})  \Brp \non \\
&=& \Abar^d M_{dc} \Blp  \om^c_{\sig} - (M^{-1})^{ca}N_{ab} p^b\Brp.
\eea
where the one form
\be
\om=dg\, g^{-1},\ \ \om^a=\Tr\, T_a dg\, g^{-1}\non
\ee
has components $\om_\tau =\gdot\, g^{-1},\   \om_{\sig}= g' g^{-1}$. We have also defined the matrices
\be
M_{ab}= h_{ab}+ D_{ab},\ \ N_{ab} = h_{ab}- D_{ab} \non.
\ee

So the constraint is now of the form
\be
\phi^i=\del x^i + \Pi^{ij} p_j
\ee
and  we can read off the bi-vector
\be
\Pi^{ab} = (h_{ac} +D_{ac} )^{-1}(h_{c}^{\ b} - D_{c}^{\ b})
\ee
which may also be written as 
\be
\Pi= \frac{1-\Ad_g}{1+\Ad_g}.
\ee

It is interesting that a closely related tensor was uncovered in the study of D-branes in WZW models\cite{Alekseev:1998mc} where it was found $D2$ branes in the WZW model wrap conjugacy classes. Although these submanifolds are topologically trivial these $D2$-branes are supported by the worldvolume $B$-field
\be
B= \frac{k}{8\pi}\Tr\, \om \frac{1+\Ad_g}{1-\Ad_g} \om.
\ee
This two form is not well defined everywhere but only on conjugacy classes. The bi-vector $\Pi$ however is well defined everywhere but not invertible.

So one can produce an topological action from the Hamiltonian \eq{GGHam}
\be
\cS_{G/G}=\int_{\Sig} \Tr \Blp \om^a\w p_a + \Pi^{ab} p_a\w p_b\Brp + \int_M \Tr \om^3
\ee
which is a Poisson sigma model twisted by an $H$-field. This is a slight fudge since the term \eq{extrapiece} is only weakly zero. Similarly, Witten showed \cite{Witten:1993xi} that the $G/G$ model is topological but one must impose the equations of motion. 

We now lift this to a membrane action, we find
\be
\cS_{G/G} =\int_M\Tr \Blp -(\om^a + \Pi^{ab}\rmp_b ) \w d\rmp_a+  f_{ab}^c \om^a\w \om^b \w p_c + f_{a}^{bc} \om^a \w p_b\w p_c + f_{abc} \om^a\w \om^b\w\om^c  \Brp. \label{GGmem}
\ee
So we see that $(H,F,Q)$ all have terms equal to the structure constants of $G$ but then there are additional terms in \eq{HFQRframes} which contain arise from \eq{GGmem} after defining $\psi^a=\om^a + \Pi^{ab}\rmp_b$ are re-organizing. These additional terms do not appear to vanish in general and are not constants. It would be interesting to understand these terms further.



\subsection{$G/H$ Gauged WZW Models}

We can also compute the Hamiltonian for the $G/H$ model in much the same way. The model is no longer topological but it is believed to be an exactly solvable CFT \cite{Gawedzki:1988nj, Gawedzki:1988hq}, the $G/H$ GKO construction. The topological terms are quite similar to the $G/G$ model however we first need to split up the Lie algebra indices $(a,b,\ldots)$ into those which are gauged ($A,B,\ldots$) and those which are not ($m,n,\ldots$). 

The action can then be written as 
\bea
I_{G/H}&=& \int_M \Tr\,(dgg^{-1})^3 + \int_{\Sigma} \Tr\Blp \half \eta^{mn} \rmp_m \w * \rmp_n + \om^m\w p_m +\om^A\w p_A + \Pi^{AB} \rmp_A \w \rmp_B   \Brp \non \\
&=&  \int_{\Sigma} \Tr \half \eta^{mn} \rmp_m \w * \rmp_n  +\int_M\Tr \Blp -(\om^a + \Pi^{ab}\rmp_b ) \w d\rmp_a+  \non \\
&&+f_{ab}^c \om^a\w \om^b \w p_c + f_{A}^{BC} \om^a \w p_b\w p_c + f_{abc} \om^a\w \om^b\w\om^c  \Brp \label{GHmem} \
\eea
with 
\be
\Pi=\frac{1-\Ad_h}{1+\Ad_h}.
\ee


\section{Future Directions}

In this work we have provided some evidence that certain generalized Wess-Zumino terms can be useful to understand geometric and non-geometric string backgrounds.  There are several problems which remain unresolved and which might help shed light on this interpretation.

Of primary interest to the author is to understand the space time field equations  in terms of the coefficients in the Roytenberg bracket. This would help in several regards, it would help to understand to what extent these coefficients are charges to be specified in a given string background and would also allow one to find new solutions of the NS supergravity equations. In the large volume limit, one can understand a WZW model as a solution of Einstein's equation which balances the spacetime curvature against the $H$ flux, it would be very interesting to find solutions which balance the $Q$-flux against the curvature. This might also shed some light on non-Abelian duality in the spirit of \cite{Klimcik:1995ux}. A key step in this direction would be to better understand the quantization conditions on the coefficients in the bracket.

We study here the bosonic string, although an interesting problem is to supersymmetrize the first order model\footnote{Some work in this direction has been done in \cite{Lindstrom:2004eh, Lindstrom:2004iw}}. One might be able to generalize the work of \cite{Bredthauer:2006hf} and find a geometric interpretation of the equivalence between the Hamiltonian and Lagrangian formulation of the $(2,2)$ sigma model. This is presumably related to the interesting recent work of Gualtieri which demonstrates the equivalence between certain holomorphic Poisson manifolds and Generalized Kahler manifolds \cite{gualtieri-2007}. Bi-vector deformations of generalized complex geometry provide an interesting interpretation of certain families of supergravity solutions \cite{Lunin:2005jy, Halmagyi:2007ft, Kulaxizi:2006zc} and it would be nice to be able to simplify this deformation procedure to find further families. The most elusive example is the NS sector of the gravity dual of the cubic deformation in $\N=4$ SYM. The difficulties here are discussed from different aspects in \cite{Halmagyi:2007ft, Kulaxizi:2006zc}, understanding the field equations in terms of $(G,\Pi)$ would presumably help understand this problem.

There are proposals in the literature for studying non-associative geometry \cite{Mathai:2004qc, Bouwknegt:2004ap} and it would interesting to embed these studies in the full dynamical sigma model. Perhaps this first order model provides an opportunity to do so. The non-associativity clearly seems to be related to the presence of $R$-flux, which as we have shown is further related to the membrane sigma model. Understanding this membrane sigma model better seems like an interesting avenue for further research.

We have written down a method for computing the action of $T$-duality on the first order sigma model by appealing to the Hamiltonian methods of \cite{Alvarez:1994wj}. However it would be very interesting to find a way to derive $T$-duality rules directly in the full sigma model at the level of the $(H,F,Q,R)$ co-efficients.

We have not yet mentioned the work of Hull and collaborators \cite{Dabholkar:2005ve, Hull:2006qs, Hull:2007jy} on non-geometric backgrounds which requires doubling the field space of the sigma model. It is not clear exactly how this is related to the first order model but it would be interesting to elucidate this. Interestingly, the beta function computation has been performed in this model \cite{Berman:2007xn}. Somewhat puzzling however is that claim in that theory that the WZW model is an example of an $R$-space \cite{Dabholkar:2005ve}, whereas here we see it as having $(H,F)$-flux. 

\vskip 10mm

\noindent {\bf {\Large Acknowledgments}}\\
I would like to thank Costas Bachas, Micha Berkooz, Peter Bouwknegt, Sameer Murthy, Giuseppe Policastro, Dmitry Roytenberg, Jan Troost and Brenno Vallilo for useful discussions. I would also like to acknowledge Ruben Minasian for collaboration at initial stages of the project. This work was supported by DSM CEA-Saclay and the grant Marie-Curie IRG 046430.

\begin{appendix}

\section{The Various Brackets}
A very nice review of derived brackets at a much more sophisticated level than that of the current work can be found in \cite{kosmannschwarzbach-2004-69}.
The Lie bracket $[.,.]$ is so well known that we deem it to be the canonical bracket, it will not carry any notation. The generalization of the Lie bracket to multi-vectors is called the Schouten bracket
\be
[.,.]_\cS : \wedge^pT\times \wedge^q T\ra \wedge^{p+q-1}T,
\ee
its explicit form is a bit cumbersome but can be found in Chapter 3 of \cite{Gualtieri:2003dx}. For example the Schouten bracket takes two bi-vectors and returns a three-vector.
On a Poisson manifold one may define the Koszul bracket 
\be
[.,.]_{\Pi}: \w^p T^*\times \w^q T^* \ra \w^{p+q-1} T^*.
\ee
On one forms this is defined as
\be
[\alpha,\beta]_\Pi=\cL_{\Pi\al} \beta - \cL_{\Pi\beta} \al +d\blp\Pi(\al,\beta) \brp
\ee
and its extension to arbitrary forms can be found for example in \cite{kosmannschwarzbach-1995}. The Koszul bracket is designed to be the analogue of the Lie and Schouten brackets acting on differential forms instead of vector fields.
The Courant bracket $[.,.]_\cC$ and the Roytenberg bracket $[.,.]_\cR$  are brackets on $T\oplus T^*$ and have been discussed in sections \ref{CBracket} and \ref{RBracket}.

\end{appendix}
\bibliographystyle{/Users/Halmagyi/utphys} \bibliography{/Users/Halmagyi/myrefs}

\end{document}